\documentclass[12pt,preprint]{emulateapj}

\usepackage{graphicx,amsmath,natbib}

\slugcomment{Submitted to ApJ}

\shortauthors{Hinkley et al.}

\begin{document}

\title{High Resolution Infrared Imaging \& Spectroscopy of the Z Canis Majoris System During Quiescence \& Outburst}
\shorttitle{High Resolution Infrared Imaging \& Spectroscopy of Z CMa}

\author{Sasha Hinkley\altaffilmark{1,12}}
\author{Lynne Hillenbrand\altaffilmark{1}}
\author{Ben R. Oppenheimer\altaffilmark{2}} 
\author{Emily Rice\altaffilmark{3,2}}
\author{Laurent Pueyo\altaffilmark{4,13}}
\author{Gautam Vasisht\altaffilmark{5}}
\author{Neil Zimmerman\altaffilmark{2,6}}
\author{Adam L. Kraus\altaffilmark{7}}
\author{Douglas Brenner\altaffilmark{2}}
\author{Charles Beichman\altaffilmark{8}}
\author{Richard Dekany\altaffilmark{9}}
\author{Jennifer E. Roberts\altaffilmark{5}}
\author{Ian R. Parry\altaffilmark{10}}
\author{Lewis C Roberts Jr.\altaffilmark{5}}
\author{Justin R. Crepp\altaffilmark{1,11}}
\author{Rick Burruss\altaffilmark{5}}
\author{J. Kent Wallace\altaffilmark{5}}
\author{Eric Cady\altaffilmark{5}}
\author{Chengxing Zhai\altaffilmark{5}}
\author{Michael Shao\altaffilmark{5}}
\author{Thomas Lockhart\altaffilmark{5}}
\author{R\'emi Soummer\altaffilmark{4}}
\author{Anand Sivaramakrishnan\altaffilmark{4}}

\altaffiltext{1}{Department of Astronomy, California Institute of Technology, 1200 E. California Blvd, MC 249-17, Pasadena, CA 91125}
\altaffiltext{2}{Astrophysics Department, American Museum of Natural History, Central Park West at 79th Street, New York, NY 10024}
\altaffiltext{3}{College of Staten Island, City University of New York, 2800 Victory Bvld, Staten Island, NY 10314}
\altaffiltext{4}{Space Telescope Science Institute, 3700 San Martin Drive, Baltimore, MD 21218}
\altaffiltext{5}{Jet Propulsion Laboratory, California Institute of Technology, 4800 Oak Grove Dr., Pasadena CA 91109}
\altaffiltext{6}{Max Planck Institute for Astronomy, Kšnigstuhl 17, 69117 Heidelberg, Germany}
\altaffiltext{7}{Harvard-Smithsonian Center for Astrophysics, 60 Garden St, Cambridge, MA, 02138}
\altaffiltext{8}{NASA Exoplanet Science Institute, California Institute of Technology, Pasadena, CA 91125}
\altaffiltext{9}{Caltech Optical Observatories, California Institute of Technology, Pasadena, CA 91125}
\altaffiltext{10} {Institute of Astronomy, University of Cambridge, Madingley Road, Cambridge CB3 0HA, UK}
\altaffiltext{11}{Current affiliation: University of Notre Dame, Dept. of Physics, 225 Nieuwland Science Hall, Notre Dame, IN 46556}
\altaffiltext{12}{NSF Fellow}
\altaffiltext{13}{Sagan Fellow}

\begin{abstract}

We present adaptive optics photometry and spectra in the $JHKL$-bands along with high spectral resolution $K$-band spectroscopy for each component of the Z Canis Majoris system.  Our high angular resolution photometry of this very young ($\lesssim$1 Myr) binary, comprised of an FU Ori object and a Herbig Ae/Be star, were gathered shortly after the 2008 outburst while our high resolution spectroscopy was gathered during a quiescent phase.  Our photometry conclusively determine that the outburst was due solely to the embedded Herbig Ae/Be member, supporting results from earlier works, and that the optically visible FU Ori component decreased slightly ($\sim$30\%) in luminosity during the same period, consistent with previous works on the variability of FU Ori type systems.   
Further, our high-resolution $K$-band spectra definitively demonstrate that the 2.294 $\mu$m CO absorption feature seen in composite spectra of the system is due solely to the FU Ori component, while a prominent CO emission feature at the same wavelength, long suspected to be associated with the innermost regions of a circumstellar accretion disk, can be assigned to the Herbig Ae/Be member. These findings are in contrast to previous analyses \citep[e.g.][]{mbd10,bmd10} of this complex system which assigned the CO emission to the FU Ori component.
\end{abstract}



\section{Introduction}
 Z Canis Majoris (hereafter ``Z CMa'') was one of the original members of the set of Ae and Be stars with nebulosity first reported in \citet{h60}, now called the Herbig Ae/Be type stars. This classification alludes not only to its early spectral type and emission lines, but also to brightness fluctuations and heavy nebulosity. At a distance of 1150 pc \citep{c74}, Z CMa is part of the CMa T1 association with a quoted age $<$1 Myr \citep{hrw78}.  Consistent with its classification and presumed age, the Z CMa system shows strong P Cyg profiles in lines of H$_{\alpha}$, H$_{\beta}$ and Fe II \citep{ctv84}, radio continuum emission, bipolar jets as well as strong infrared excess \citep{pmr89}.  Initial efforts to image the emitting source at high resolution revealed a $\sim$100 mas elongation \citep{lh87}, suggesting a nearly edge-on disk-like structure.  Shortly thereafter, \citet{kbs89} suggested that the extended emission was probably not due to a disk, but rather due to a cooler, close companion retaining luminosity from its pre-main sequence contraction.

\begin{figure*}[ht]
\center
\resizebox{1.10\hsize}{!}{\includegraphics[angle=0]{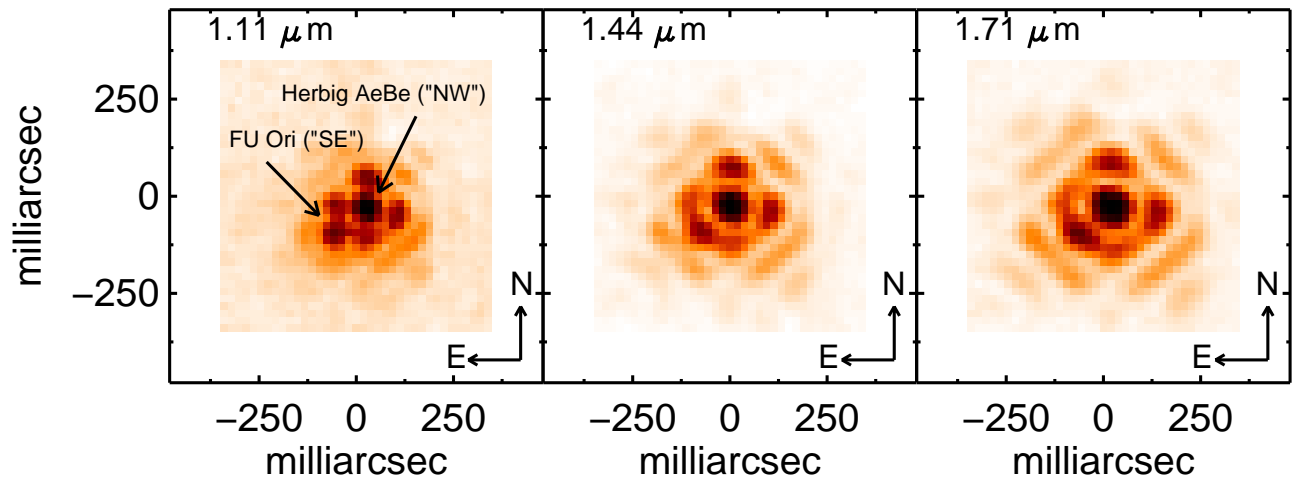}}
  \caption{Three panels showing the Z CMa system shortly after its 2008 outburst at 1.11, 1.44, and 1.71 $\mu$m, obtained using the ``Project 1640'' Integral Field Spectrograph at Palomar Observatory \citep{hoz11}. The two members of this system  are separated by $\sim$1.5 Palomar near-infrared diffraction widths. At longer wavelengths, the Northwest member, often referred to as a ``Herbig Ae/Be'' object, dominates the emission over the Southeast member, the ``FU Orionis'' object. }
\label{zcma} 
\end{figure*}

At about the same time, \citet{hkh89} suggested the Z CMa system had many characteristics of FU Ori-like systems \citep[see e.g.][]{hk96}, including a bright accretion disk with a high rate of  accretion, and signatures of powerful winds. Progress on reconciling the Herbig Ae/Be nature of the star with its now apparent FU Ori-like characteristics was achieved when evidence was uncovered through speckle imaging for {\it two} objects comprising the system \citep{kbg91}.  These measurements were later verified by visible direct imaging \citep{tbb95} revealing two clearly defined point sources. Such results provided support to the hypothesis that a significant fraction of FU Ori type systems may be binary systems \citep{bb92}, which is indeed the case for the FU Ori system itself \citep[][Pueyo et al. 2012]{wah04,ra04}.  

Synthesizing several decades of observations, a consistent picture of the Z CMa system has now emerged of a binary system comprised of a FU Ori-like object which dominates the optical emission, and a Herbig Ae/Be object, which dominates the infrared \citep{wcs93,vbt04}. The terminology used for each component in the system in past literature has been somewhat ambiguous given the constantly evolving physical understanding of this remarkable system. As such, in this work we have chosen to abandon the ``primary'' and ``secondary'' nomenclature that pervades the literature, and instead use the terms ``Z CMa Southeast (SE)'' to refer to the FU Ori component and ``Z CMa Northwest (NW)'', referring the the Herbig Ae/Be component.  \citet{kbg91} show that the SE component dominates the emission of the NW component blueward of 2$\mu$m, and vice versa for wavelengths longer than 2 $\mu$m.   

Support for the basic physical picture of the system has been verified repeatedly over the past ten years: i.e. the identification that the SE component is a $\sim$few $M_\odot$ object \citep{vbt04} responsible for the jet/outflow phenomena \citep{mgm02}, while the NW component has considerable evidence supporting its Herbig Ae/Be nature including emission lines and a compact, massive envelope \citep{wcs93}.  
This system experienced a significant outburst in 2008 \citep[e.g.][]{ga09} lasting $\sim$1.5 years, and was photometrically monitored extensively at X-ray to visible wavelengths \citep{mbd10}.  Indeed, using spectropolarimetry of the system, \cite{shs10} claimed the outburst was associated with the Herbig Ae/Be component of the system.  Further, while dual-imaging polarimetry has typically been reserved for young circumstellar disk systems \citep[e.g.][]{obh08,hos09}, recently \citet{cmj12} have used this technique to identify the jets associated with {\it each} member of this system. 

\begin{figure*}[ht]
\center
\resizebox{1.0\hsize}{!}{\includegraphics[angle=0]{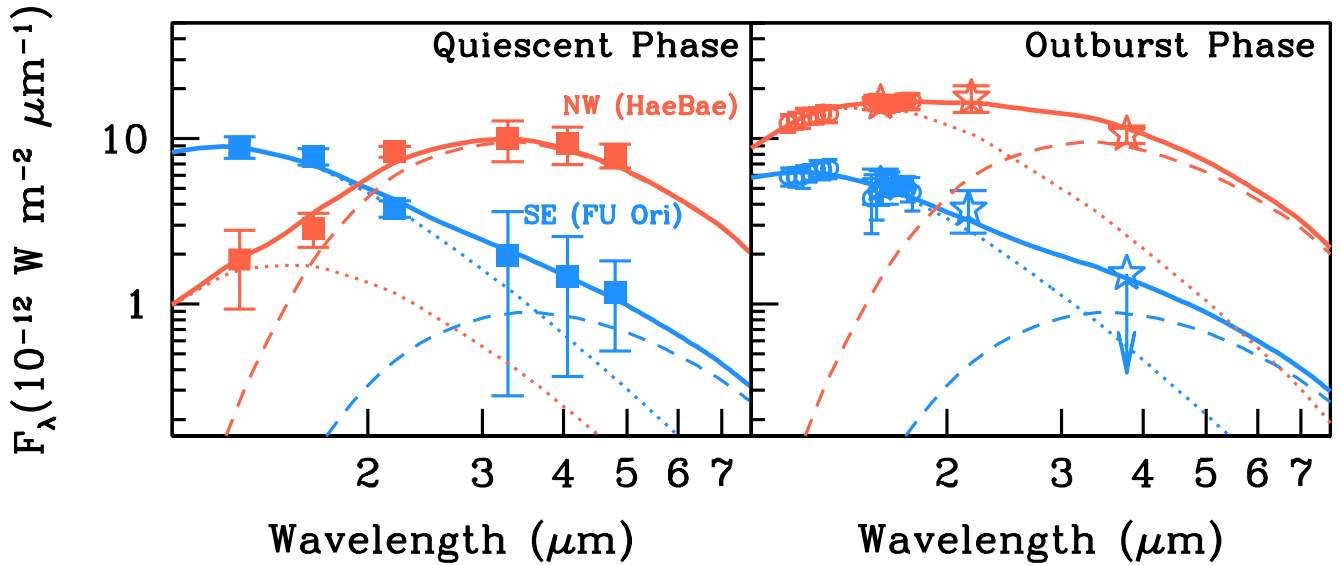}}
  \caption{
{\it Left:} Spectral Energy Distributions of each component of the Z CMa system in the quiescent state taken from \cite{kbg91}.  The NW (Herbig Ae/Be) component is shown with the red points while the photometry for the SE member (FU Ori) is shown in blue.   {\it Right:} In the same color scheme, $1.06-1.76 \mu$m outburst spectra obtained from the Palomar Project 1640 spectrograph on 2009 March 17 (open circles), as well as $HKL$ Keck/NIRC2 points obtained on 2008 December 21 indicated by star-shaped points.  Each solid curve represents two-component models comprised of the sum of two reddened blackbodies shown by dotted and dashed lines.  The 2008 outburst of this system is conclusively due to the order-of-magnitude brightness increase by the Herbig Ae/Be component.  
}
\label{bothstates} 
\end{figure*}

Near-infrared spectroscopy of the Z CMa system  has been scarce over the past 30 years.  Indeed, as \citet{aag09} point out, only \citet{lo97} and \citet{vbt04} have provided previous near-infrared spectra.  In this paper, we present near-IR spectra and/or photometry in the $JHKL$ bands for both components during the 2008 outburst, resolved using high angular resolution imaging \citep{oh09}, as well as archival photometry covering the same bandpasses \citep{kbg91} for comparison. 
Our adaptive optics (``AO'') measurements directly demonstrate and provide confirmation of  previous work \citep[e.g.][]{shs10, mbd10} that suggested the outburst is due to the Herbig Ae/Be component, although the exact mechanism of the brightening of this member remains unknown. Moreover, we present high-resolution $K$-band spectra of each member of the binary obtained in 2006 during the quiescent phase. Table~\ref{obstable} provides a list of our observations. 

\section{Observations and Data Processing}
Our high angular resolution observations of Z CMa were taken using Palomar/P1640 and Keck/NIRC2 AO during outburst. Our AO-assisted $K$-band spectroscopy was taken with Keck/NIRSPEC in quiescence. Table~\ref{obstable} provides a summary.  

\subsection{$JHKL$-band Spectra \& Photometry During Outburst}
We imaged the Z CMa system in its high state on UT 2009 March 17 using ``Project 1640'' \citep{hoz11,obb12} on the 200-in Hale Telescope at Palomar Observatory. Project 1640 is a coronagraph integrated with an integral field spectrograph (IFS).  The IFS+Coronagraph package is mounted on the Palomar AO system \citep{dbp98}, which in turn is mounted at the Cassegrain focus of the Hale Telescope.  The coronagraph is an Apodized-Pupil Lyot coronagraph \citep[APLC;][]{skm01,s05}, having a 370 mas diameter (5.37$\lambda/D$ at $H$-band) focal plane mask. The IFS, is a microlens-based imaging spectrograph which can simultaneously obtain $\sim$40,000 spectra across our $3.8^{\prime\prime}\times3.8^{\prime\prime}$ field of view. Each microlens subtends 19.2 mas on the sky and a dispersing prism provides a spectral resolution ($\lambda/\Delta\lambda$)$\sim$45. The observing wavelengths spanned the $J$ and $H$-bands (1.06$\mu$m - 1.76$\mu$m at the time of these observations).
Early examples of spectrophotometry and astrometry from this project can be found in \citet{hob10}, \citet{zoh10}, \citet{hmo11}, \citet{rrb12}, and \citet{phv12}. While an IFS clearly aids in spectral characterization, it can also improve sensitivity \citep{cpb11,pcv12} through the suppression of quasi-static speckle noise in the direct image which will limit high contrast observations \citep[][]{hos07}.   This system is also equipped with aperture masking interferometry capabilities \citep[e.g.][]{hci11,ki12} for the characterization of systems with very small angular separations, although this technique was not employed in this study. 

Three individual wavelength channels are shown in Figure~\ref{zcma}. To fully resolve both members of the system, the target was observed $\sim$1 arcsec away from the coronagraphic mask. 

Point Spread Function (``PSF'') fitting photometry was performed in each of the 23 wavelength channels using a calibration star, HD 112196 (F8V, $V$=7.01), obtained  on the same night with similar signal-to-noise and observed at similar airmass.  The 100 mas angular separation of the two Z CMa components (Figure~\ref{zcma}) 
is only slightly larger than the 70 mas $H$-band diffraction limit of the Palomar Hale Telescope. Given this fact, fitting the empirical calibration PSF star to each member was performed simultaneously.  Each channel of the Z CMa images were oversampled by a factor of five, and the calibration PSF was fit to each Z CMa component to determine peak brightness and centroid positions.     
Once these two best-fit PSFs were determined, each was subsequently subtracted from their corresponding Z CMa component to determine the post-subtraction residual flux.  This residual flux is the dominant term in the calculation of the uncertainties in the $J$ and $H$-band spectra.

To avoid saturation on such a bright target, the outburst data obtained with NIRC2 at Keck Observatory on UT 2008 December 21 utilized the $J_{\rm cont}$, $H_{\rm cont}$, $K_{\rm cont}$, and PAH narrow filters with central wavelengths 1.213 $\mu$m, 1.580 $\mu$m, 2.271 $\mu$m, and 3.290 $\mu$m, respectively. The flux values in these narrower filters were then scaled to corresponding broadband $JHKL$ wavelengths using the ratio of bandpasses between the two sets. The Z CMa system was observed using three-point dithering, and airmass corrected using the calibration stars listed in Table~\ref{obstable}. 

The left panel of Figure~\ref{bothstates} shows the photometry of each component in the quiescent state collected from \citet{kbg91} using speckle interferometric observations.  The right panel of Figure~\ref{bothstates} shows our $J$ and $H$-band $R$$\sim$45 spectra from Palomar, as well as our $HKL$-band photometry obtained at Keck for both members of the system taken from the 2008-2009 outburst observations.   

\begin{deluxetable*}{llcclcc}
\tabletypesize{\scriptsize}
\tablecaption{Table of Observations}
\tablewidth{0pt}
\tablehead{ 
\colhead{Object} & 
\colhead{Star Type} & 
\colhead{Date}  &   
\colhead{Z CMa State} & 
\colhead{Observatory \& Instrument} & 
\colhead{Wavelengths ($\mu$m)} &
\colhead{Mode}
}
\startdata
Z CMa          & Target        & 2006 Dec 17      & Quiescence    & Keck: NIRSPEC             & 2.10-2.13, 2.29-2.32  & Spectroscopy \\ 
S Mon          & Calibrator    & 2006 Dec 17      &              -       & Keck: NIRSPEC             & 2.10-2.13, 2.29-2.32  & Spectroscopy \\ 
Z CMa          & Target        & 2008 Dec  21     &  Outburst        & Keck: NIRC2                   & 1.57-3.32                                & Imaging \\         
HD 53455     & Calibrator    & 2008 Dec  21     &              -       & Keck: NIRC2                   & 1.57-3.32                                 & Imaging \\         
HD 75898     & Calibrator    & 2008 Dec  21     &           -          & Keck: NIRC2                   & 1.57-3.32                                & Imaging \\         
Z CMa          & Target        & 2009 Mar  17     &   Outburst       & Palomar: Project 1640      & 1.06-1.76                                & Spectro-photometry  \\
HD 112196    & Calibrator   & 2009 Mar  17     &        -              & Palomar: Project 1640      & 1.06-1.76                               & Spectro-photometry  
\enddata
\label{obstable}
\end{deluxetable*}

\subsection{High Resolution $K$-band spectroscopy During Quiescence}
Spatially resolved $K$-band spectra of the two Z CMa components were obtained with the NIRSPEC spectrograph mounted behind the adaptive optics system on the W. M. Keck II telescope \citep{was00} on UT 2006 December 17.   The $K$-band setup provided wavelength coverage across the CO bandhead, the Br $\gamma$ line, a He I line,  and a Mg I and Al I metallic line region, but the spectral coverage is not continuous.   The $R$=$\lambda/{\Delta\lambda}$$\sim$25,000 observations were obtained at two position angles:  60 degrees (along the jet axis) and 120 degrees (along the projected semi-major axis of the binary).   The slit was 2.26$^{\prime\prime}$ long and data were taken using two different widths: 0.068$^{\prime\prime}$ and 0.027$^{\prime\prime}$.  MCDS readout mode was used with 30s and 60s integration times with five co-adds.  Because of the complexity of the field, in addition to dithered pairs of spectra taken in an ABBA nod pattern, off-target sky images were also obtained.   Telluric correction was achieved using the rapidly rotating early type star S Mon and further calibration was derived from internal arc line lamps and flat field lamps.  
Figure~\ref{nirspec} shows the flux from two echelle orders from NIRSPEC, spanning $2.10 - 2.13 \mu$m and $2.29 - 2.32 \mu$m. 

\section{Results}

\subsection{Confirmation of the Outbursting Component}\label{p1640photometry}
The quiescent flux measurements taken from \citet{kbg91} and shown in Figure~\ref{bothstates} reveal that shortward of $\sim$2 $\mu$m, the overall system flux is dominated by the SE component (blue curve in Figure~\ref{bothstates}).  Longward of $\sim$2 $\mu$m, the NW component (the Herbig Ae/Be star shown by the red points) dominates.  Further, the NW component has increased in brightness during outburst by an order-of-magnitude over the quiescent values.  At the same time, we find a $\sim$30\% flux decline between the quiescent and outburst states for the SE component (the FU Ori star), shown in blue.  The high angular resolution afforded by the Keck and Palomar AO systems allows us to confirm the claim \citep{mbd10,bmd10,shs10} that the NW component is responsible for the ensemble system's increase in brightness. 

 
\subsection{Two-component Model Fits to Data}
Using the photometry obtained at Keck and Palomar, as well as the work of \citet{kbg91}, we are able to satisfactorily fit two-component, reddened, blackbody models to the quiescent and outburst Spectral Energy Distributions for each member of the Z CMa system shown in Figure~\ref{bothstates}.   For the NW component, our model that is physically plausible for such a luminous star is made up of two blackbodies of temperatures 8500 and 1100K behind $A_V$=10. For the SE member, we find a physical match using components of 5500 and 900K behind $A_V$=5.2.  The reddening uses a optical total-to-selective extinction ratio, $R_v$=3.1, and the \citet{m90} extinction curve.  Although we have chosen to refrain from assigning a great deal of physical meaning to this two-component model, we refer to these components as ``photosphere'' and ``disk''.  In each of the quiescent and outburst phases, the disk blackbody temperatures, normalizations, and reddening values are identical. At the same time, the normalization of the fit to the photospheric component of the NW member has increased by an order of magnitude from quiescence to outburst, while for the SE component, it has been reduced by 30\% between the two stages.  The inherent variability of the Z CMa system is reflected in these changing photosphere normalizations.  

Figure~\ref{bothstates} shows that in addition to the order-of-magnitude brightness increase of the NW component, the peak luminosity of this member has shifted bluewards, indicating either a greater luminosity of the photospheric component, or a significant decrease in the $A_V$ value.  However, decreasing the $A_V$ value immediately requires unphysical values for the photospheric temperature. Specifically, $A_V$=2 for the NW component implies an unphysical best fit photosphere temperature of $\sim$1300K. The 5500K photospheric temperature for the SE member agrees well with typical FU Ori objects that show CO absorption features (See Section~\ref{cosection}) as described, for example, in \citet{cpm91}.  Meanwhile the disk temperature for the NW member agrees well with values quoted in \citet{shs10}.  Lastly, this $\sim$30\% decrease in flux for the SE component over $\sim$20 years is remarkably similar to the optical decay reported in \citet{clm05} for the archetypal system FU Ori.

\subsection{CO Features Observed in Quiescence}\label{cosection}
\begin{figure}[ht]
\center
\resizebox{1.0\hsize}{!}{\includegraphics[angle=0]{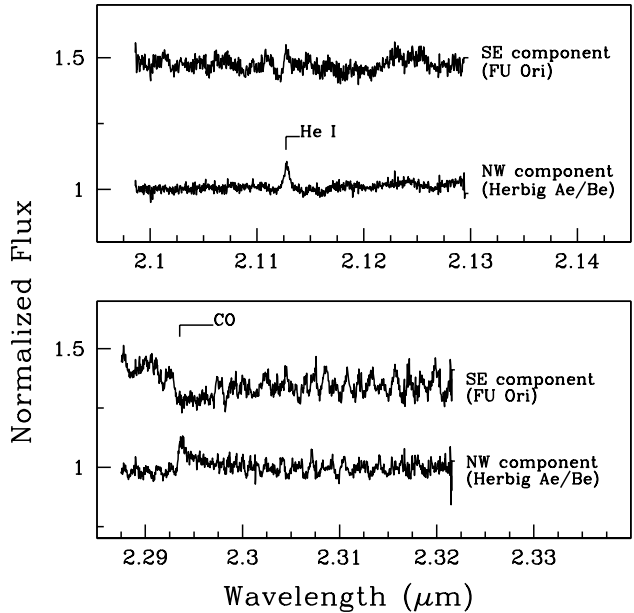}}
  \caption{Keck $K$-band spectra from the NIRSPEC spectrograph for both the SE and NW components of Z CMa obtained while the system was in quiescence in 2006 December 17.  The Herbig Ae/Be component shows a prominent CO emission feature located at 2.293 $\mu$m, while the FU Ori component shows a significant absorption at this wavelength. }
\label{nirspec} 
\end{figure}

The NIRSPEC spectra shown in Figure~\ref{nirspec} 
capture the prominent He I emission line at 2.113 $\mu$m as well as the CO bandhead at 2.294 $\mu$m.  CO emission in this wavelength range occurs in $\sim$20\% of luminous young stellar systems \citep[e.g.][]{c89}, and probes the innermost regions of the circumstellar accretion disks \citep{ncc08}.  
 

Our Keck high angular resolution observations allow us to determine that the SE component (the FU Ori star) possesses a significant CO bandhead absorption at 2.294 $\mu$m.   Several FU Ori objects such as FU Ori itself, V1057 Cyg, and V1515 Cyg, show significant absorption features due to this CO band head  \citep[e.g.][]{ra97}.  It has been noted that Z CMa possesses rather weak CO absorption at 2.294 $\mu$m relative to other FU Ori stars, attributed to mixing of the signals between the two components due to inadequate angular resolution \citep{hk96}.  We can check this hypothesis with our high angular resolution observations.  Our clear detection of the absorption band head at 2.294 $\mu$m in the SE component confirms the claim of \citet{vbt04} that the absorption is most likely arising from this component. While \citet{lo97} present evidence for the CO band head using medium resolution $K$-band spectra (see especially their Figure 1), the spectral resolution of their observations was not sufficient to ascribe these CO bandhead features to one component or the other. Indeed, these authors attempt a decomposition of the spectrum by choosing a linear analytic function for the NW component, which forces deviations from this linear form to be due to the SE component.   


At the same time, Figure~\ref{nirspec} shows that the NW companion (the Herbig Ae/Be) shows an emission feature at the band head location.  \citet{mbd10} and \citet{bmd10} note the presence of this CO overtone line in emission during outburst in their spatially unresolved observations of the Z CMa system, but assume that this arises from the SE component (the FU Ori member). Our high spectral resolution observations {\it clearly demonstrate the CO emission arises from the NW component}. \citet{cpm91} discuss at length the conditions under which an object may exhibit CO emission during an outburst phase through the increased irradiation of the atmosphere of the circumstellar disk. 





\section{Conclusions}
We present high angular resolution near-infrared observations of the Z Canis Majoris system during its 2008 outburst as well as prior to this during a quiescent phase.  The $JHKL$-band outburst photometry conclusively determines 1) that the brightening is due solely to the embedded Herbig Ae/Be member, confirming results from earlier works, and 2) that the optically visible FU Ori component has actually experienced slightly declining brightness between the quiescent and outburst stages. Thus, this substantial brightness increase of the Herbig Ae/Be component relative to its quiescent state is responsible for the continuum brightness doubling of the ensemble system. Further, the Keck high-resolution $K$-band spectra obtained during a quiescent phase allow us to conclusively determine that the 2.294 $\mu$m CO absorption feature seen in composite spectra of the system is due solely to the FU Ori component.  In addition however, these $K$-band observations show a prominent CO emission feature in the Herbig Ae/Be member which likely dilutes the strength of the corresponding absorption feature in the SE component when the images of each component are blended with lower angular resolution observations. This result stands in contrast to assumptions in other works about the source of the CO emission \citep[e.g.][]{bmd10,mbd10}. 


\acknowledgments 
{\it Acknowledgments:} 
We thank the anonymous referee for several helpful suggestions.  This work was performed in part under contract with 
the California Institute of Technology (Caltech) funded by NASA 
through the Sagan Fellowship Program. A portion of this work is or was supported by the National Science Foundation under Grant Numbers AST-1203023, AST-0804417, 0334916, 0215793,  0520822, and 1245018. A portion of the research in this paper was carried out at the Jet Propulsion Laboratory, California Institute of Technology, under a contract with the National Aeronautics and Space Administration and was funded by internal Research and Technology Development funds. The authors would like to thank Pat Hartigan for his help obtaining the Keck/NIRSPEC observations.  A portion of the research in this paper was carried out at the Jet Propulsion Laboratory, California Institute of Technology, under a contract with the National Aeronautics and Space Administration.   Some of the data presented herein were obtained at the W.M. Keck Observatory, which is operated as a scientific partnership among the California Institute of Technology, the University of California and NASA. The Observatory was made possible by the generous financial support of the W.M. Keck Foundation.  



\end{document}